# The formation and assembly of a typical star-forming galaxy at redshift 3.

Daniel P. Stark[1], Mark Swinbank[2], Richard S Ellis[1], Simon Dye[3], Ian Smail[2],

& Johan Richard[1]

*1 Department of Astronomy, Caltech, Pasadena CA 91125 USA*

*[2] Institute for Computational Cosmology, Durham University, South Road, Durham DH1 3LE, UK*

*[3]School of Physics & Astronomy, Cardiff University CF24 3AA, UK*

Recent studies of galaxies ~2-3 Gyr after the Big Bang have revealed large, turbulent, rotating structures[1-2]. The existence of well-ordered rotation in galaxies during this peak epoch of cosmic star formation may suggest that gas accretion through cold streams is likely to be the dominant mode by which most star-forming galaxies at high redshift grow since major mergers of galaxies can completely disrupt the observed velocity fields. But poor spatial resolution and sensitivity have hampered this interpretation, limiting the study to the largest and most luminous galaxies, which may have fundamentally different modes of assembly than more typical galaxies. Here we report observations of a typical star forming galaxy at z=3.07 with a linear resolution of ~100 parsec. We find a well-ordered compact source in which molecular gas is being converted efficiently into stars, likely assembling a spheroidal bulge and disk similar to those seen in spiral galaxies at the present day.

Resolved imaging spectroscopy provides a powerful route to probing the internal properties of young galaxies. Using adaptive optics, which corrects for the effects of atmospheric motion, internal velocity maps have been acquired for a small number of galaxies at redshifts of z=2-3 (2-3 billion years after the Big Bang) on scales of 0.15 arcsec (~1.3 kiloparsec)[1-3]. These early results do not yet reveal a clear picture. The largest galaxies studied show evidence for systematic rotation, consistent with forming spiral disks[1,2]. However, smaller more typical galaxies observed at the same epoch show chaotic velocity fields consistent with more primitive systems[3]. The interpretation is still limited by poor spatial resolution: typical galaxies at such redshifts have angular sizes of order 0.25 arcsec and so, even with adaptive optics, the velocity field is poorly sampled. Moreover, the emission lines used to trace the velocity field are not uniformly bright across the galaxy and co-adding pixels to improve the signal comes at the expense of a degradation in resolution.

An effective way to resolve these issues is to use *gravitational lensing* which magnifies the images of distant galaxies[4-7] allowing a much more detailed view than otherwise possible. To demonstrate the advances possible, we have combined laser guide star adaptive optics with gravitational lensing to study a star-forming galaxy (MACS J2135-0102[8]) at redshift 3.075 (seen 2.1 Gyr after the Big Bang). Detailed modelling of this system reveals that this galaxy is magnified in area 28±3 times (and ~8 times along its major axis) by the combination of a foreground galaxy and galaxy cluster[9] and that its properties are typical of Lyman-break galaxies at this redshift.

To map the redshifted [OIII] 4959,5007 and Hβ 4861 Å emission lines, we employed the OSIRIS integral field spectrograph[10] on the 10m Keck II telescope. Figure

1 shows a colour image of the target constructed from the optical Hubble Space Telescope image and the OSIRIS infrared spectroscopic data; clearly the two datasets are well-matched in resolution. Utilising adaptive optics, together with the gravitational magnification, the image plane FWHM corresponds to a physical scale of just ~120 parsecs in the source plane. Not only does this yield a ~5 times improvement in linear resolution in sampling the velocity field compared to earlier work, but the signal can be coadded in fainter regions whilst still maintaining an improved resolution compared to that achievable for unlensed sources.

In order to interpret our observations, we take into account the magnification caused by the foreground lens and reconstruct the resolved galaxy in the *source plane*, making use of the detailed mass model[9], which defines the mapping from each location in the *image plane*. Figure 2 shows the resulting intensity and velocity maps where binning has typically been undertaken with source plane pixels of 0.02 arcsec, corresponding to a sampling of 150 pc. The rest-frame UV continuum distribution reveals an elongated morphology with two components. Component 1 is the most highly magnified (a factor 28±3 times) and gives rise to the ring-like structure in the image-plane. Component 2 extends just outside the caustic giving rise to a lower magnification (a factor 8.0±0.9 times). These large magnifications enable us to study the internal properties of J2135-0102 in unprecedented detail. Fitting both the continuum and [O III] emission line distributions, we derive a half-light radius of 750 parsecs, a factor of five smaller than the large galaxies studied by many earlier workers[1,2] and much closer to the typical size of z~3 LBGs[11].

The most significant result revealed in Figure 2 is the well-sampled, regular, bi-symmetric velocity field revealed by the [O III] emission line measures which, together with the line widths which peak at the center of the galaxy, suggest a rotating system.

But could the observed velocity pattern arise from a merging pair whose projected velocities are smoothed into an apparent rotation curve as a result of inadequate spatial resolution? This ambiguity, arising from the poorly-sampled data typically available at high-redshift, has plagued the interpretation of the first resolved data from distant galaxies[1-4], thereby precluding a robust test of the hypothesis that a large fraction of distant star-forming galaxies are undergoing major mergers[12]. With our considerably improved resolution, this ambiguity is greatly reduced. The well-sampled data for J2135-0102 reveals a velocity gradient spanning many independent resolution elements *even internally with component 1*. Moreover, the projected velocity of component 2 is consistent with an extrapolation of the gradient across component 1, which would be highly coincidental if the components were separate systems. We thus conclude that J2135-0102 it is much more likely to be a single system undergoing rotation.

Examining the velocity field in more detail, we extract a one-dimensional rotation curve along the major axis of the velocity field (at a position angle of 10º clockwise from the $\Delta y$-axis). Using an *arctan* function to model the shape of the rotation curve, the observations suggest a circular velocity of $v_C \sin i = 55 \pm 7$ km s$^{-1}$ where $i$ is the inclination angle. We estimate $i = 55º \pm 10º$ from the surface photometry, implying a circular velocity of 67 km s$^{-1}$. As described above, component 2 shares the rotation, affecting only a minor $7 \pm 3$ km s$^{-1}$ adjustment coincident with a 25 km s$^{-1}$ rise in the line width. Using the rotation curve we derive a dynamical mass of $\sim 2 \times 10^9$ solar masses within a radius of 1.8 kpc. Thus, compared with the stellar mass of $6 \pm 2 \times 10^9$ solar masses[13] (assuming a Salpeter initial mass function), the central region appears to be baryon-dominated. While J2135-0102 certainly represents the most convincing example of a rotating galaxy, significant random motions are present. The central [O III] velocity dispersion $\sigma_0 = 54 \pm 4$ km s$^{-1}$ is comparable to the rotational velocity and the ratio $v/\sigma_0 = 1.2 \pm 0.1$ is consistent with previous, poorer-sampled, observations of similar-sized galaxies at this redshift[3]. Any disk present in J2135-0102 is thus most likely at an early stage of assembly (see Supplementary Information for further discussion).

Our OSIRIS data also yields the 2-D distribution of the nebular line Hβ, a valuable tracer of recent (<20 Myr) star formation. The integrated star formation rate derived from the Hβ flux is 40 ± 5 $M_\odot$ yr$^{-1}$ (uncorrected for extinction)[14], consistent with that derived from the 24μm flux[13]. The star formation rate density of 4.4 ± 0.5 solar masses yr$^{-1}$ kpc$^{-2}$ is typical of that derived for a larger sample of z~2-3 UV-selected galaxies[15,16] and comparable to that in local starbursts where activity is usually in compact, circumnuclear disks. The high star formation rates observed in these galaxies result in significant stellar winds and supernovae ejecta, which are thought to contribute to the vigorous galactic scale outflows have been observed in LBGs[17]. The blue-shifted UV-interstellar absorption lines (Δv = -400 ± 100 km s$^{-1}$)[8] are consistent with this interpretation.

Importantly, we also show that the Hβ velocity field confirms the overall rotation. However, both the Hβ and CO emission[13] are not co-spatial with the center of the galaxy, but appear to peak within component 2. Moreover, the CO velocity and line width are in good agreement with those seen in [O III] for this component suggesting that almost all of the cold gas (which acts a reservoir for future star formation) may reside in a region less than 1 kpc across. We utilize additional spatially-resolved ESO VLT SINFONI observations of [O II], [O III] and Hβ (see Supplementary Information) to measure the $R_{23}$ abundance index[18,19], finding a metallicity of ~0.9 solar metallicity for both components.

These key findings allow us to determine the nature of J2135-0102. The interplay between the dynamics, star-formation and mass of the gas reservoir in high redshift star-forming galaxies is one of the primary science drivers for the next generation facilities such as the Atacama Large Millimeter Array (ALMA). Since J2135-0102 is currently the only Lyman Break Galaxy for which high resolution kinematic *and* cold gas properties are available, we are uniquely able to dynamically constrain the gas mass and compare this with that implied using local calibrations, thereby testing whether star formation occurs in a mode similar to that seen in local spirals or appropriate to that in more extreme systems. The ratio $\alpha$, between the CO luminosity and the molecular gas mass, has been used to distinguish the mode of star-formation[20]. For local spirals, $\alpha \sim 5$, consistent with star formation occurring in discrete giant molecular clouds with $10^5$-$10^6$ solar masses. In luminous infrared galaxies, where star formation occurs more vigorously in a single reservoir with $10^9$ solar masses, $\alpha$ is close to unity[21]. $\alpha$ has yet to be measured at high redshift for all but the most extreme populations[22]. By requiring that the gas mass must be less than the dynamical mass in J2135-0102 we find $\alpha < 0.8$. This implies that the gas lies in an extensive intercloud medium rather than in discrete, less massive, gravitationally-bound clouds[21]. The lower conversion factor identified here suggests that the gas masses of rest-UV selected galaxies at $z \sim 2$-3 may be lower than previously thought[23].

1. Genzel, R., Tacconi, L., Eisenhauer, F. et al, The rapid formation of a large rotating disk galaxy three billion years after the Big Bang. *Nature*, **442**, 786-789 (2006).

2. Forster-Schreiber, N.M., Genzel, R., Eisenhauer, F. et al, SINFONI integral field spectroscopy of z~2 UV-selected galaxies. *Astrophys. J.*, **645**, 1062- 1075 (2006).

3. Law, D., Steidel, C.C., Erb, D. K., Larkin, J. E., Pettini, M., Shapley, A. & Wright, S., Integral field spectroscopy of high redshift star-forming galaxies with laser-guide star adaptive optics, *Astrophys. J.*, **669**, 929-946 (2007).

'**Supplementary Information** accompanies the paper on **www.nature.com/nature**.'

Acknowledgements: We thank Jim Lyke for assistance with the Keck observators and acknowledge useful discussions with Richard Bower, Kristen Coppin, Matthew Lehnert, Reinhard Genzel, Dawn Erb, David Law, Alice Shapley, Adrian Jenkins, Paolo Salucci and Tom Theuns. The OSIRIS data were obtained at the W.M. Keck Observatory which is operated as a scientific partnership among the California Institute of Technology, the University of California and NASA. The observatory was made possible by the generous financial support of the W.M. Keck Foundation. The SINFONI data are based on observations made with the ESO Telescopes at the Paranal Observatories. AMS and IRS/RSE acknowledge financial support from STFC and the Royal Society respectively.

The authors declare no competing financial interests.

Correspondence and requests for materials should be addressed to DPS (e-mail: dps@astro.caltech.edu).

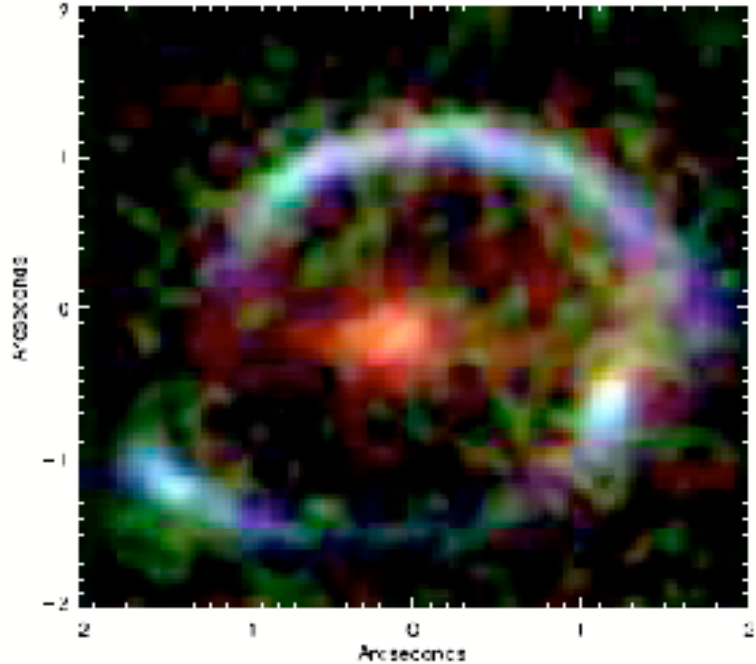

**Figure 1**: Colour image of J2135-0102 – a z=3.07 galaxy magnified 28 ± 3 times into a near-complete Einstein ring[8]. This image combines a Hubble Space Telescope $V_{606}$ image (blue) with Keck Laser Guide Star Adaptive Optics assisted near-infrared images in [O III] 5007 Å (green) and broad-band K (red). The foreground lensing galaxy at z=0.7 is the resolved red source at the center of the image. The large extent of the ring compared to the image plane resolution of our observations (0.13 arcsec) enables us to map the variation in spectral and dynamical properties of the background galaxy in fine detail.

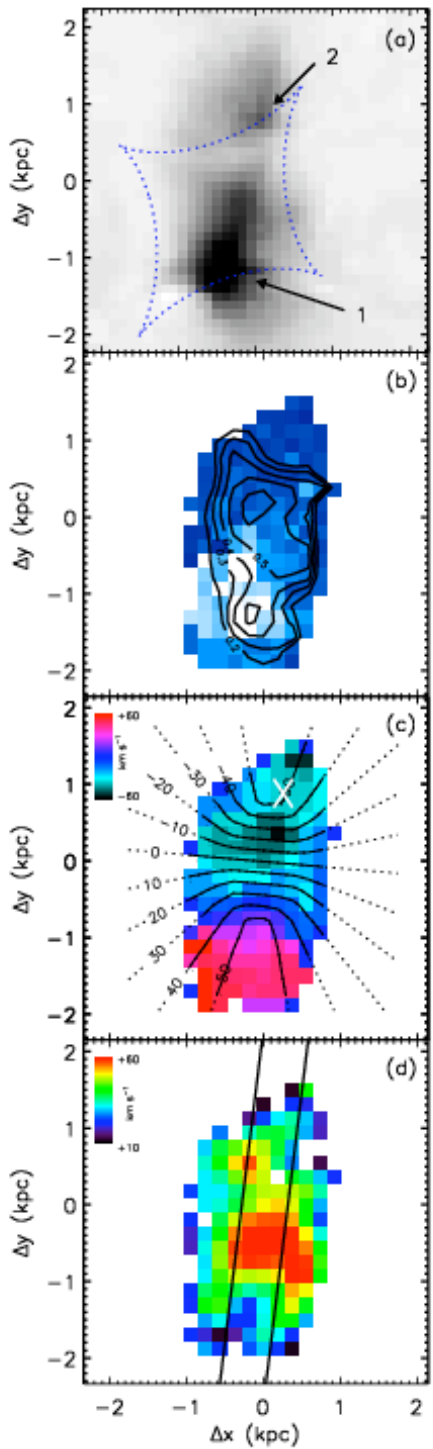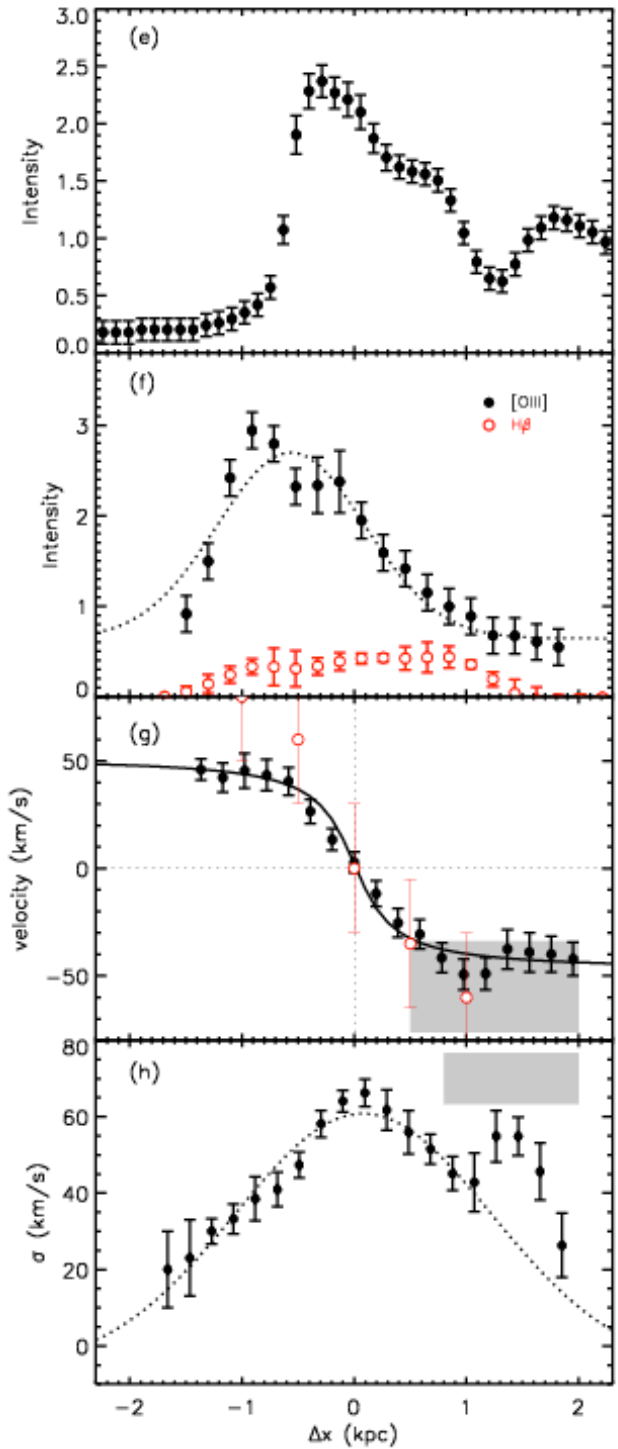

**Figure 2**: The internal dynamics of J2135-0102, reconstructed in the source plane using the well-constrained lens model[12]. Regions with the largest and smallest magnifications have source plane resolutions from 0.02" to 0.04", corresponding to physical scales of ~150-300pc. For clarity we have resampled to a single pixel scale of 0.02". Each pixel thus represents a nearly independent measurement. **Left (top to bottom):** (a) Reconstructed HST $V_{606}$ image of J2135-0102 showing the morphology of the rest-frame UV stellar continuum. Two major components within the galaxy are labelled as 1 and 2. The lens model caustic is overlaid (blue dashed line); regions close to the caustic have the highest magnification. (b) Distribution of [O III] 5007 Å emission (blue image) compared with that from Hβ (contours), which traces the star formation rate with contours varying linearly from 0.2 to 0.6 $M_\odot$/yr/pixel. (c) Velocity field derived from the [O III] emission. Overlaid contours indicate the best-fit disk model (see Supplementary Information). "X" marks the centre of the cold molecular gas[16]. (d) Distribution of the rest-frame standard deviation, σ, of the [O III] emission line. **Right (top to bottom):** (e) $V_{606}$ surface brightness profile. The extraction is performed along the long-axis of the galaxy (-10° from the Δy-axis) with Δx=0 corresponding to a location of (0,-0.2) on the left-hand side. (f) Spatial profiles and corresponding 1σ Poisson uncertainties of [O III] and Hβ. (g) Major axis rotation curve for [O III] (black points) and corresponding 1σ uncertainties verified with lower significance using Hβ (red points) with best-fitting disk model overlaid (solid line). (h) Profile of the [O III] emission line standard deviation and corresponding 1σ uncertainties showing a maximum of σ = 54 ± 4 km s$^{-1}$.

Shaded regions in panels (g) and (h) show the spatial and kinematic location of the molecular CO emission[13] and the minor influence of component 2 on the overall dynamics.

## Supplementary Information

**Details of the Keck/OSIRIS Observations:** The strongly-lensed z=3.07 galaxy, J2135-0102[1] was observed with OSIRIS[2] in conjunction with the Laser Guide Star Adaptive Optics system on the Keck II 10 metre telescope on UT September 4-5 2007 in photometric conditions and an optical seeing of 0.5 arcsec FWHM. The resulting Strehl is 25% and the encircled energy is 50% within a radius of 0.10 arcseconds. In order to map the nebular emission lines of [O III] λλ4959, 5007 Å and Hβ 4861 Å, we used the narrow band Kn1 filter which provides a wavelength coverage from 1.955 to 2.055μm at a spectral resolution of $\lambda/\Delta\lambda \sim 3400$. When deriving emission line widths, we subtract the instrumental profile, whose FWHM is $5.9 \times 10^{-4}$ μm, in quadrature. As the target only partially fills the OSIRIS integral field unit (IFU) using the 0.1 arcsec lenslet scale, observations were conducted using a ABBA sequence with a 3.2 arcsec East-West chop onto adjacent sky, keeping the target within the field of view. The total integration time comprised 20 sub-exposures of 750 seconds. Individual exposures were reduced using the OSIRIS data reduction package. From each resulting datacube, a continuum image was constructed and the centroid of the foreground lensing galaxy measured. The final datacube was constructed by aligning the sub-exposures and combining using a 3σ clip to reject cosmic rays. Flux calibration was obtained by observing the R=14.2 `tip tilt' star prior to each observation.

**Gravitational Lens Modelling:** To interpret our data we must correct for the gravitational magnification and reconstruct the galaxy in the source plane. We first assign accurate world coordinates to each spectral pixel using the known coordinates of the foreground lens and the position angle of the IFU. The source plane is subsequently reconstructed with the lens model held fixed to the parameterization determined using the high signal/noise HST data[3] taking care to account for the PSF of the OSIRIS observations.

We note that the variable magnification across J2135-0102 leads to non-uniform resolution in the source plane; however we take this into consideration during the source-plane reconstruction of the data cube by utilizing an adaptive source plane grid[4]. Regions in the image plane that are more highly magnified map to correspondingly smaller source plane pixels. The pixel size in each region of the source plane is selected by ray-tracing the image of the PSF into the source plane and ensuring that the derived resolution is finer than the pixel size. Hence each pixel in the source plane represents a nearly independent measurement of the intensity and velocity field. The pixel size ranges from 0.02 arcseconds in the regions of high magnification to 0.04 arcseconds in the regions of lower magnification, corresponding to physical scales of ~150 to 300 pc at z=3.07.

To evaluate the uncertainty on the magnification factor and on the image reconstruction, we also create a family of acceptable lens models. These models are constructed by varying the four main parameters of the lens model, namely the normalization of the dark matter halo density distribution of the lens galaxy, the elongation of the halo (the major / minor axis ratio), the orientation of the major axis, and the external shear provided by the cluster, such that the $\Delta\chi^2$ value from the best fit varies by $\pm 1\sigma$. Ray tracing each of these models and re-deriving the source-plane morphology and velocity field produces only minor changes in the dynamics, typically the various mappings produce a maximum shift of ~0.01 arcseconds compared to the best-fit solution and only a 10% uncertainty in the derived magnification (Supplementary Figure 1). All of the derived quantities in this paper include this uncertainty.

**Analysis of Emission Line Maps:** The distribution of velocities and line widths across J2135-0102 was computed by fitting the [O III] λ5007 emission line in the spectral direction at each spatial location with a flat continuum plus Gaussian emission line profile using a $\chi^2$ minimization procedure which takes into account the greater noise level close to atmospheric OH emission. We demanded a minimum signal to noise ratio of 5 to detect the emission line, and when this criterion is met, we determine the centroid, flux and line width of the best fit. In regions where the fit failed to detect a line, we average the surrounding 2 × 2 pixels and attempt the fit again. To derive error bars on each parameter we perturb the best fit such that the $\Delta\chi^2$ value from the best fit varies by ±1σ.

Star formation rates are derived from the Hβ line fluxes following the Kennicutt calibration[5] which assumes a Salpeter IMF ranging from 0.1 to 100 solar masses and case B recombination. We rule out a significant contribution to the recombination flux from an active galactic nucleus (which would bias our star formation rate estimates) from the absence of bright CIV emission[1] and bright mid-IR continuum[6]. The integrated Hβ line flux implies a star formation rate of 40 solar masses / year. This value likely underestimates the star formation rate of J2135-0102 since it does not account for dust. If we adopt an extinction correction of A(Hβ)~1 mag, estimated by taking into account the reddening suggested from the broadband SED[6], the intrinsic star formation rate of J2135-0102 is closer to 100 solar masses / year. This is still consistent with the mid-infrared star formation rate[6] (60 solar masses / year) and is identical to that predicted from the extinction corrected rest-UV continuum[1].

**Rotation vs. Merger Interpretation:** Throughout the article, we argue that the most consistent explanation of our velocity and line width information (Figure 2) follows if the galaxy has well-ordered rotation. Here we discuss whether other interpretations are possible. In particular, are the data consistent with a merging system or a bipolar outflow?

Both observations and simulations have shown that gaseous disks are almost always completely destroyed immediately after an equal mass merger[7-9]. Hence, if J2135-0102 is in the intermediate or late stages of a merger, we would expect the kinematics to be strongly disturbed. Instead, we see a relatively smooth rotation profile across the long axis of the galaxy with line widths peaking near the galaxy center, as predicted in simple models of rotating disks. We note that simulations of equal mass mergers at high redshift suggest that with inadequate angular resolution (~0.5-1.0" FWHM), the velocity distortions induced by a merger event may be smeared out giving the appearance of an undisturbed velocity field[9]; however, on the ~150 pc scales probed by the OSIRIS observations of J2135-0102, the simulations suggest that such substructure should be readily apparent in the velocity field. Of course, the residual angular momentum from a galaxy merger can lead to the formation of a disk[10] in as little as ~100 Myr, making it difficult to rule out the possibility that the ordered motions seen in J2135-0102 were initially generated in a major merger.

Alternatively, it is possible that the system is in the early stages of a merger. In this scenario, component 2 is seen in projection with a line-of-sight velocity coincident with the extrapolation of the rotation curve of component 1. Assuming that all velocity vectors are equally probable for the component 2, we derive a <1% probability that the relative spatial position and radial velocity of the component would lie within 10 km/s from component 1 and ±15 degrees from its location on the major axis defined in Figure 2. We therefore view this interpretation as unlikely.

We also consider whether the dynamics could be explained due to an outflow from the central source. While we do expect strong superwinds in J2135-0102, the outflow velocity suggested by the offset between the interstellar absorption features (400±100 km/s)[1] is several times larger than the velocity gradient seen in [OIII] and H$\beta$ (~100 km/s). Furthermore, for a typical starburst galaxy in the local universe only 3-4% of the H$\alpha$ luminosity arises from shock-heated gas[11]. This emission also has a much lower surface brightness and is likely difficult to recover at high-redshift. Recalling that H$\beta$ shows the same velocity structure as [O III], these observations suggest that the H$\beta$ and [OIII] luminosity are dominated by radiative processes associated with star formation and not shocks generated in the outflow.

**The Rotating Disk Model:** In order further evaluate the nature of the velocity field, we explore exponential disk models. We construct simplified two dimensional velocity fields assuming that the velocity follows an arctan function[12]. We allow the peak rotational velocity ($v_c$), turn-over radius ($r_{peak}$), inclination angle (*i*), and position angle ($\theta$) to vary. For each disk model, we construct a datacube with the same pixel scale as our observations and add a Gaussian emission line with line centroid and width reflecting the local dynamics. At each pixel we also add noise appropriate for our data and then refit the velocity field using the same $\chi^2$ minimization line fitting code. We find that the best fit model has a position angle +10º clockwise from the $\Delta$y-axis, $v_c$ *sin i* = 54.7 km/s, $r_{peak}$=300pc (Supplementary Figure 2). In Fig. 2c we show the two dimensional velocity field of J2135-0102 and overlay the contours from the best model fit. As can be seen, the model provides a reasonable fit to the data, with the contours tracing the global velocity field, as shown by the one dimensional extracted rotation curve (solid circles) and corresponding model (solid line). The velocity dispersion map and one-dimensional cut are also well-matched to the model (Fig. 2e).

While this simple disk model provides a reasonably good fit to the data, it is worth noting that we are *not* suggesting J2135-0102 is analogous to the cold, disk galaxies seen in the local universe. Indeed, there are numerous small-scale deviations from the model, most notably at the location of the gas-rich component (at Δx=1.5 kpc on the right-hand side of Fig. 2), where the extracted rotation curve shows an 8 km/s rise and the velocity dispersion shows evidence of a secondary peak. Such substructure might be explained via a disk instability which has led to rapid star formation in a subregion within J2135-0102 and thereby caused a disruption of the local dynamics. Disturbances to the velocity field and line widths are in fact predicted in simulations of unstable disks[13-15]. We also note that our simple disk model fails to explain the 500 pc offset in the peak of the line width distribution from our adopted galaxy center. This is perhaps not surprising considering that in the timescale required to assemble the observed stellar mass given a star formation rate of 60 solar masses yr$^{-1}$ (~100 Myr), a test particle at a radius of 1 kpc will have only completed ~1-2 orbits of the galaxy at the observed rotational velocity; hence it is likely still in a fairly unrelaxed state.

As discussed in the main paper, the line widths are comparable to the inclination-corrected rotational velocity of the galaxy ($v_c / \sigma_0$ = 1.2 ± 0.10) suggesting considerable random motions. This could arise in two ways[16]. First, as gas is accreted, disk viscosity may convert gravitational potential energy into random motions. This heating may, however, inhibit the efficient dissipative processes necessary for the formation of the clumps seen in J2135-0102. Perhaps instead, random motions were generated *after* star formation was initiated as a result of the stellar winds and outflows present in this galaxy[17]. Which of these two scenarios plays the dominant heating role can be determined with knowledge of the relative distribution of stars and ionized gas. In the latter scenario, random motions are generated after the burst of star formation causing the ionized material to be extended relative to the distribution of stars. In the former case, the heating occurs before the star formation episode, leading to an equally extended distribution of stars and ionized gas. High-resolution near-infrared imaging will soon provide the necessary constraints on the stellar distribution of J2135-0102, enabling this test to be conducted.

**Metallicity Constraints:** Additional data covering the redshifted [O II] 3727 Å emission line doublet was obtained using the SINFONI IFU and its associated Laser Guide Star Adaptive Optics system on the 8.2 metre ESO Very Large Telescope UT4 (Yepun). A 3,600 sec (on source) exposure was secured in photometric conditions in 0.6 arcsec seeing using the same tip-tilt reference star as in the OSIRIS observations. These data offer a 3.2 × 3.2 arcsec field with a sampling of 0.10 arcsec pixel$^{-1}$. The data have a lower spectral resolution ($\lambda/\Delta\lambda \sim 1500$ at 1.5μm) and (due to the shorter exposure time) poorer signal-to-noise than the OSIRIS data, but cover a wider wavelength range, $\lambda = 1.4 - 2.4$ μm. The observations were reduced in a similar manner to that described above using the ESOREX pipeline.

This data allows us to examine the chemical composition using the $R_{23}$ index[18,19] derived from the ratio of [O II], [O III] and Hβ emission. Although each line comprising this index is clearly detected over the entire source, the signal-to-noise is insufficient for a measurement in each spatial pixel. However, the data do allow us to bin the spectra in component 1 and 2 separately for which we measure $R_{23} = 4.7 \pm 0.8$ and $3.8 \pm 0.4$, respectively.

In deriving metallicities from the $R_{23}$ index, uncertainty arises from the double-valued nature of the calibration. We adopt the calibration based on the upper-branch of this relation, yielding $12 + \log(O/H) = 8.6$ for the entire galaxy, for two reasons. First, this gives a result that is closer to the median metallicity of a large sample of z~2 UV-selected galaxies measured using the [NII]6583/Hα calibration (12+log(O/H) ~ 8.7)[20] and a smaller sample of z~3 LBGs measured assuming the upper branch of the $R_{23}$ calibration (12+log(O/H) ~ 8.6)[21]. Secondly, our choice is also more consistent with the stellar mass – metallicity relation established for z~2 LBGs[20] given the inferred stellar mass for J2135-0102[6]. Adopting the upper branch calibration and estimating the excitation parameter via the ratio of [O III] and [O II], we find that the oxygen-phase abundance is ~0.9 solar (12 + log(O/H) ~ 8.6) in both components. The lack of a pronounced metallicity gradient in J2135-0102 is consistent with the interpretation that the galaxy is still in a fairly unrelaxed state, having only been actively forming stars for little over a dynamical time.

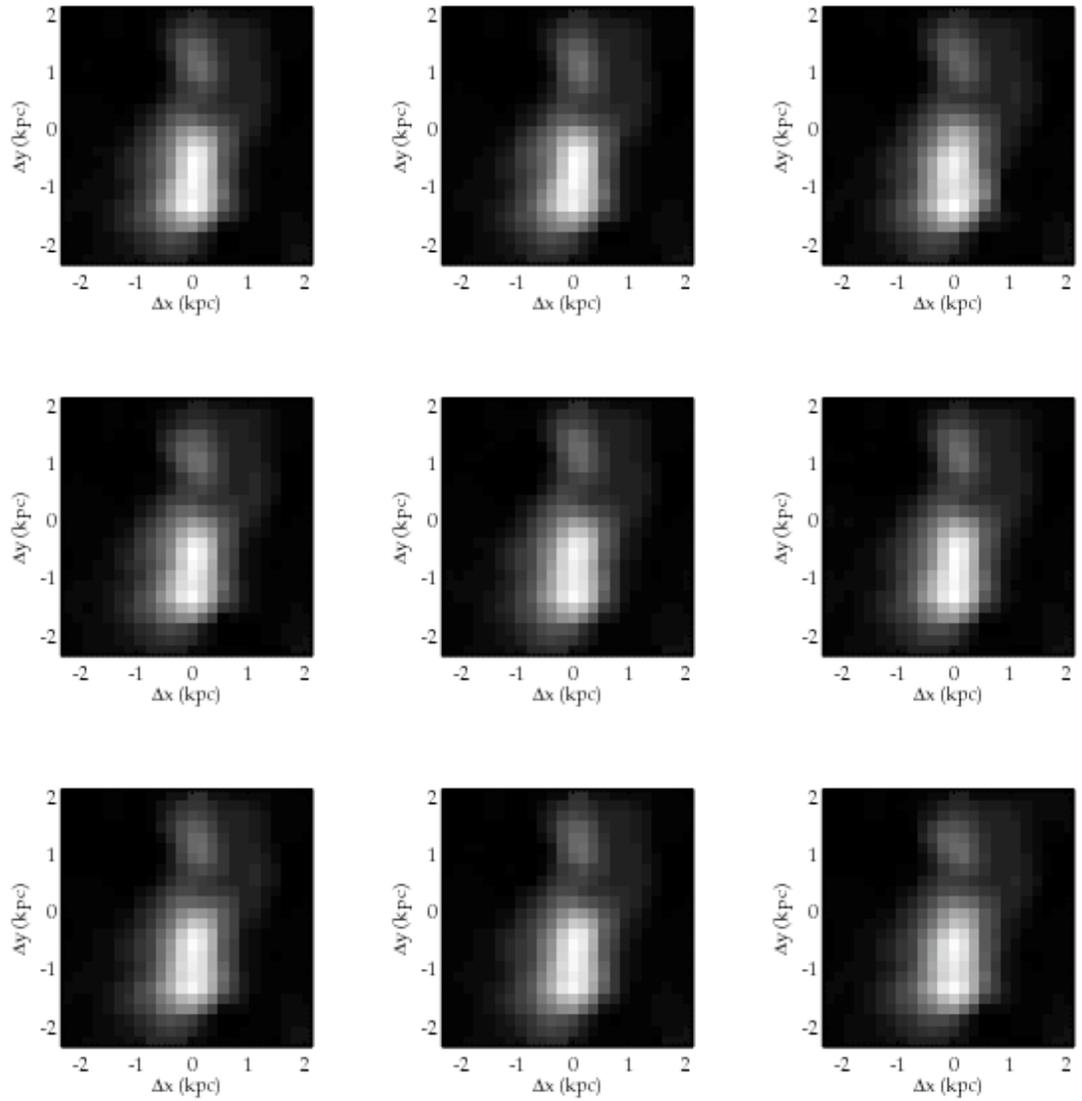

**Supplementary Figure 1**: Uncertainties in source plane reconstruction. The nine different models are constructed by varying the normalization, elongation, orientation, and external cluster shear of the lens model (see text for details) by ±1σ from their best-fit values. Ray tracing each of these models and re-deriving the source-plane morphology and velocity field produces only minor changes in the dynamics, typically the various mappings produce a maximum shift of ~0.01

arcseconds compared to the best-fit solution and only a 10% uncertainty in the derived magnification. For consistency with Figure 2, the adaptive source plane grid (with pixel sizes varying between 0.02 and 0.04 arcseconds) is resampled onto a uniform grid with 0.02 arcsecond pixels.

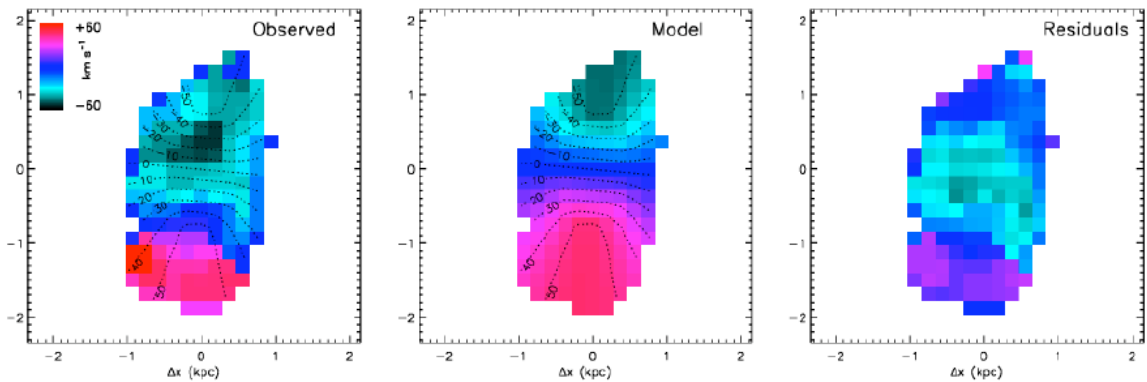

**Supplementary Figure 2**: Best-fitting disk model and corresponding residuals. The data are fit using a simple arctan function[12] allowing the peak rotational velocity, turn-over radius, inclination angle, and position angle to vary. The model that minimizes $\chi^2$ has $\theta$ = +10° clockwise from the $\Delta$y-axis, $v_c$ $\sin i$ = 54.7 km/s, $i$ =55°, $r_{peak}$=300pc. The best-fitting disk model is able to reproduce the velocity shear with reasonably low residuals over most of the velocity field. Regions with larger residuals (<20 km s$^{-1}$) may potentially reflect substructure within disk, as predicted for unstable disks undergoing rapid star formation[13-15].

**Supplementary References:**